# Main Manuscript for

## Early volatile depletion on planetesimals inferred from C-S systematics of iron meteorite parent bodies

**Marc M. Hirschmann[1,*], Edwin A. Bergin[2], Geoff A. Blake[3], Fred J. Ciesla[4], Jie Li[5]**

[1] Department of Earth and Environmental Sciences, University of Minnesota, Minneapolis, MN 55455.
[2] Department of Astronomy, University of Michigan, Ann Arbor, MI 48109.
[3] Division of Geological & Planetary Sciences, California Institute of Technology, Pasadena, CA 91125.
[4] Department of Geophysical Sciences and Chicago Center for Cosmochemistry, University of Chicago, Chicago, IL 60637.
[5] Department of Earth and Environmental Sciences, University of Michigan, Ann Arbor, MI 48109.
*Corresponding author: M.M. Hirschmann mmh@umn.edu
.

**Author Contributions:** This work is the product of collaborative discussions among all the coauthors over a period of five years. M.M. Hirschmann compiled meteorite compositional data, performed the thermodynamic and planetesimal degassing calculations, and wrote the initial manuscript draft. Additional contributions include connections to nebular and protoplanetary disk processes and exoplanets (E.A. Bergin and G.A. Blake), planetesimal differentiation and planetary accretion processes (F.J. Ciesla), and the behavior of S and C during planetesimal core-formation processes (J. Li). All of the coauthors edited and improved manuscript drafts. Author order after the lead is alphabetical.

**Competing Interest Statement:** None.
**Classification:** Major Classification: Earth, Atmospheric, and Planetary Sciences. Minor classification: Astronomy

**Revised Manuscript, February 22, 2021**

**Abstract**

During the formation of terrestrial planets, volatile loss may occur through nebular processing, planetesimal differentiation, and planetary accretion. We investigate iron meteorites as an archive of volatile loss during planetesimal processing. The carbon contents of the parent bodies of magmatic iron meteorites are reconstructed by thermodynamic modelling. Calculated solid/molten alloy partitioning of C increases greatly with liquid S concentration and inferred parent body C concentrations range from 0.0004 to 0.11 wt.%. Parent bodies fall into 2 compositional clusters characterized by cores with medium, and low C/S. Both of these require significant planetesimal degassing, as metamorphic devolatilization on chondrite-like precursors is insufficient to account for their C depletions. Planetesimal core formation models, ranging from closed system extraction to degassing of a wholly molten body, show that significant open system silicate melting and volatile loss is required to match medium and low C/S parent body core compositions. Greater depletion in C relative to S is the hallmark of silicate degassing, indicating that parent body core compositions record processes that affect composite silicate/iron planetesimals. Degassing of bare cores stripped of their silicate mantles would deplete S with negligible C loss, and could not account for inferred parent body core compositions. Devolatilization during small-body differentiation is thus a key process in shaping the volatile inventory of terrestrial planets derived from planetesimals and planetary embryos.

**Significance Statement**

*Habitable rocky worlds require a supply of essential volatile elements (C,H,N,S). These are plentiful in early solar systems, but depleted during processes leading to planet formation. Here new evidence for loss during differentiation of small precursor bodies or planetesimals is derived from iron meteorites, which are samples of planetesimal cores. Reconstruction of the C and S contents of planetesimal cores indicates severe C depletions compared to inferred original planetesimal compositions. Modeling of depletion processes shows that preferential loss of C compared to S is transferred to cores during differentiation. Iron meteorites preserve evidence of a key devolatilization stage in the formation of habitable planets and suggest pervasive carbon loss is likely associated with the birth of terrestrial worlds.*

## Introduction

Major volatiles (H, C, N, and S) are inherently plentiful in the interstellar medium and abundant in primitive carbonaceous chondrites (1, 2), but are scarce in terrestrial planets, which gained most of their mass from the inner parts of the solar nebula (3, 4). Formation of volatile-poor planets from a volatile-rich protoplanetary disk is a result of processes in the solar nebula, in accretion of precursor solids, and in interior differentiation. Addition of volatiles to nascent planets varies during accretion as protoplanetary systems become dynamically excited, contributing material originating from different heliocentric distances (3) and with different thermal histories. Much of this mass arrives in larger bodies (planetesimals or planetary embryos) that differentiated soon after formation (5). Key uncertainties include the nebular history of bulk materials that contributed volatiles to the rocky planets and how that affected their volatile cargos (6), and how planetesimal and planet formation influenced volatile distributions in accreted parent bodies.

Processes responsible for volatile deficits in terrestrial planets (7, 8) can occur either in the *nebular*, *planetesimal*, or *planetary* environment. *Nebular* volatile depletion could result from chemical interactions between nebular gas and dust, chondrule formation, or the accretion of thermally processed solids (9-11), perhaps owing to the hotter conditions prevailing closer to the protosun (4). Li et al. (6) argue that the comparatively small C inventory of the bulk Earth requires that nebular materials experienced significant early (<1 Ma) heating, before the "soot line" moved inward of 1 AU. *Planetesimal* processes involve loss to space during differentiation or processing of intermediate-sized bodies of 10s to 100s of km in diameter (e.g.,12, 13). *Planetary* loss processes occur on large (1000s of km diameter) bodies (14, 15) in which gravity plays an appreciable role - including loss from impacts (16). The sum of these is an important determinant for whether terrestrial planets form with volatiles sufficient for habitability but not so great as to become ocean worlds (17) or greenhouse hothouses (18).

A key goal in the study of exoplanets and of young stellar systems is predicting environments and processes that could lead to habitable planets, including development of models that account for the distribution, acquisition, and loss of key volatile elements. Astronomical studies can reveal the architecture of other solar systems (19), the compositions of observable exoplanet atmospheres (20 and references therein), and the dust and volatile gas structure and composition of protoplanetary disks (21 and references therein), including interactions of the disk with gas- or ice-giant protoplanets. However, only limited astronomical observations can be made about conversion of disk materials (gas, dust, and pebbles) to planets in other solar systems. To understand this conversion, we must necessarily rely on planetesimals and their remnants (meteorites) as records of the processes that occurred. In this paper, we focus on volatile loss during planetesimal differentiation by examining evidence chiefly from iron meteorites. We note that ephemeral metal enrichments in white dwarf atmospheres confirm that differentiated planetesimals are common around other stars (22), and that our findings apply to how materials would have been processed during the assembly of other planetary systems.

In classic oligarchic growth models of planetary origin, planets and embryos grow from accretion of planetesimals with characteristic radii of 10s to a few 100s of km (3). In pebble accretion models of terrestrial planet formation, the fraction of planetesimals in accreting material varies with time and protoplanetary mass (23), but still remains significant. Thus, for understanding volatile delivery to growing planets, an important question is whether the volatile inventory of accreting planetesimals (or larger objects) remained similar to that of primitive materials, typically taken to be comparable to chondritic meteorites[*], or had diminished significantly from prior differentiation.

Achondritic meteorites[†] are fragments of differentiated planetesimals and provide direct evidence of processes on small bodies. Evidence for volatile loss on silicate achondritic parent bodies comes from elemental concentrations and from isotopes (24-27). However, the best-studied silicate achondritic suites, such as the HEDs and angrites, are igneous crustal rocks (28), and their compositions may not reflect average major volatile contents of their parent bodies. Volatile loss could have been locally enhanced by the igneous activity that produced the planetesimal crusts (29).

Iron meteorites offer an additional record of volatile processing in planetesimals. Many, known as "magmatic" irons, originated as metallic cores of planetesimals

---

[*] *Chondrites* are meteorites that contain chondrules and are considered "primitive" in that they did not undergo enough thermal processing to produce phase separation, such as removal of molten metal or silicate.
[†] *Achondrites* are meteorites that have experienced sufficient thermal processing to destroy the chondrules that are characteristic of more primitive chondritic meteorites. These include both silicate achondrites as well as iron meteorites.

(30) and potentially record volatile depletions in their parent planetesimals at the time of alloy-silicate separation. Iron meteorites contain measurable amounts both major (S,C,N) and moderately volatile (Ge, Ga) elements and represent the cores of at least 50 parent bodies (31). Thus, known parent body cores are likely survivors from a population of planetesimals that were mostly incorporated into larger bodies and planets. Additionally, isotopic evidence links iron meteorites with both carbonaceous (CC) and non-carbonaceous (NC) chondrites (32), thereby correlating the differentiated planetesimals to their primitive chondritic heritage.

Here we address the problem of planetesimal volatile loss by focusing on carbon and sulfur, two siderophile[‡] volatile elements that give important clues to the degassing history of metallic cores recorded iron meteorites and thereby their parent planetesimals. We begin by examination of C-S systematics in different classes of chondrites. Although chondritic parent bodies formed later than most parent bodies of iron meteorites (33), they provide the best available guide to undifferentiated materials in the early solar system. Their isotopic kinships to iron meteorites (32) suggest that they derive from similar, though not necessarily identical, reservoirs and so they provide a basis for comparison to those estimated for parent body cores. They also reveal devolatilization processes associated with planetesimal metamorphism. We then examine iron meteorite groups and reconstruct the compositions of their respective parent cores. Finally, we consider a spectrum of simple planetesimal core-formation scenarios and model the resulting C and S distributions. Comparison of these to reconstructed parent core

[‡] “siderophile” elements are those that tend to concentrate in metallic alloys.

C and S places new constraints on the magnitude of degassing occurring from planetesimal interiors.

**S and C in parent cores and other cosmochemical objects**

To explore variations in C and S concentrations in planetesimals, we employ a *log* C/S versus *log* C plot (34) (Fig. 1). Processes leading to C enrichment or depletion without S variation produce diagonal trends on this graph, whereas independent S variations at constant C yield vertical trends.

**C-S variations in primitive planetesimals**

Primitive planetesimals have not undergone differentiation to a metallic core and silicate mantle (±crust), and are represented in meteorite collections by chondrites and by primitive achondrites (28). Average C and S concentrations for chondrite groups (see SI Appendix, Table S1) show coherent variation in *log* C/S versus *log* C along a diagonal trend that reflects significant depletions in C with more modest reductions in S concentration (Fig. 1A). CI, CM, CR and TL carbonaceous chondrite groups comprise the C-rich end of the trend while the strongly depleted CK carbonaceous and ordinary chondrites are the most C-poor. Intermediate concentrations are found in enstatite chondrites and CO, CV, CB carbonaceous chondrites.

For the ordinary (H, L, LL) chondrite parent bodies, average concentrations taken from the compilation of Wasson and Kallemeyn (35) obscure considerable variations as a function of petrologic type[§], with higher types having lower C and C/S (Fig. 1B), which must reflect metamorphic processing on their parent planetesimals. Variations seen between carbonaceous chondrite parent bodies reflect a similar dependence on petrologic type, but are evident mainly between groups from different parent bodies, which makes it difficult to separate the relative effects of nebular processes from those associated with planetesimal processes.

Primitive achondrites, which have experienced metamorphism sufficient to destroy textural evidence of chondrules, but not undergone wholesale silicate/alloy separation (28), are slightly displaced from the chondrite trend, with acapulcoites having compositions similar to chondrites of high petrologic type, but winonaites more similar to chondrites with intermediate petrologic type (e.g., H3, CV3, etc.).

**C-S variations in parent body cores**

Our goal is to use iron meteorite groups to estimate the composition of their molten parent cores prior to crystallization. Defining S and C concentrations in parent cores presents multiple challenges, owing partly to segregation of S and C in accessory phases and coarse heterogeneous textures that impede representative "average" bulk analyses (37) and partly to compositional variation within groups produced by extended fractional crystallization (38, 39). Further, it is commonly inferred that selective destruction of S-rich differentiates causes S contents observed in irons to be biased to low concentrations not representative of bulk

---

[§] *Petrologic type* is a measure of thermal and aqueous processing of the parent body, usually between types 1 and 6, with higher numeric values corresponding to greater temperatures (36)

parent core compositions (40-42). For these reasons, estimates of S and C contents of parent body cores can conflict. Recognizing that each parent core must have unique mean S and C concentrations, we here adopt the strategy of considering multiple values where assessments diverge. This exercise will illustrate the extent to which the conclusions of this paper depend on the different estimates.

Analyzed bulk sulfur concentrations in iron meteorites generally range from 0 to 2 wt.% (43), but many iron meteorite parent bodies are thought to be more S-rich (38-40, 42). Reconstruction of initial parent liquid S concentration is feasible for the magmatic iron meteorite groups, which show coherent fractional crystallization trends for trace elements (38, 39, 42, 44-48).

Because S has strong effects on liquid/solid alloy partition coefficients of trace elements (38, 40), bulk S contents of parent body cores can in theory be estimated from elemental modeling for those groups that show coherent differentiation trends. Such exercises indicate S-rich initial liquids ranging from 0.2 to 17 wt.% (SI Appendix, Table S2), but resulting quantitative estimates show large differences depending on the methodology adopted. For example, Chabot (39) estimated 17 and 12 wt.% S, respectively, in parental magma for the IIAB and IIIAB clans, whereas Wasson et al. (47) estimated 6 wt.% S for IIAB and Wasson (38) 2 wt.% for IIIAB parental liquids. The differences arise from different physical models for melt-solid segregation, with Wasson (38) and Wasson et al. (47) emphasizing the role of trapped liquid. We consider both sets of estimates (SI Appendix Table S2).

Evaluation of C concentrations in iron meteorite groups is elaborated in the *SI Appendix*. Low concentrations of C, ranging from 0.006-0.15 wt.% (SI Appendix Table S2), are a salient feature of the magmatic iron meteorites and are chiefly lower than even the most depleted chondrite groups (L 0.09 wt.%; CK 0.07 wt.%; SI Appendix Table S1). Because C is siderophile (49), cores formed by close-system segregation of molten metal from chondrite-like planetesimals would be expected to have greater C.

An important question is whether the low C concentrations inferred for parent body cores reflect the compositions of average solids crystallized from planetesimal cores, or if, similar to S, they are biased to low concentrations owing to systematic loss of samples formed from C-enriched liquid. However, unlike S, C has modest solubility in taenite (the FeNi alloy that crystallizes from melt, also known as austenite)(50). If magmatic iron meteorites represent cumulates, formation of C-rich liquids by fractional crystallization should be recorded by conjugate C enrichments in cumulates formed from more evolved liquids. But within meteorite groups, C concentrations do not increase with Ni (51, 52), the chief major element indicator of differentiation. This observation lead Goldstein et al. (37) to conclude that such C-rich liquids do not develop during fractional crystallization of planetesimal cores. They noted that the C-Ni systematics of iron groups contrast with correlations between P and Ni, which signify evolution of P-rich differentiates. The absence of C-Ni correlations suggests that estimates of C inferred from samples of an iron group may approximate original parent body core C concentrations, even if the particular meteorites crystallized from fractionated liquids.

A maximum estimate for the C contents of initial parent core liquids can be calculated by assuming that the igneous mineralogy of iron meteorites represents pure accumulations of crystallized alloy, without any trapped liquid (See SI Appendix Text). For the magmatic irons, the C content of parent liquids from which these cumulates precipitated can be found by matching the activity of C in solid Fe-Ni-C metal with that of a coexisting metal-sulfide-carbide liquid, using the liquid S contents inferred above. For this calculation, we employ a thermodynamic model of Fe-Ni-C taenite (50) and of Fe-C-S liquid (53). We neglect the effects of Ni on the liquid C activity liquid, which would diminish slightly the calculated liquid C content. (54, 55). To the extent that some iron meteorites contain precipitated trapped liquid (38), rather than consisting of purely cumulus alloy, this calculation will overestimate the C of the calculated liquid for cases in which C behaves as an incompatible element, which applies when S contents are below ~10 wt.%, and underestimate it for very S-rich liquids (See SI Appendix Text for further discussion).

For low S-liquids, calculated C concentrations are modestly greater than inferred cumulate compositions, but for high S liquids, they are lower than the C of the corresponding solid meteorites (Fig 2). The latter effect arises because S strongly enhances the C activity coefficient in Fe(Ni)-C-S liquids (53), as has been well-documented in experimental studies of graphite solubility in sulfide liquids (55-58). Consequently, during planetesimal core crystallization C behaves as a modestly incompatible element in liquids with <10 wt.% S and a compatible element for more S rich liquids (Fig. S1). This contrasts with the common assumption that C is a strongly incompatible element during solidification of planetesimal cores (37, 59) and provides an explanation for why C-enrichment is not recorded in the C-Ni systematics of individual parent bodies (37). Therefore, from both an empirical and theoretical perspective, the low C contents of magmatic iron meteorites are indicative of low C concentrations in parent cores, and not systematically biased by under-sampling of putative C-rich late crystallized products.

The calculated C concentrations of parental liquids for each parent body are given in SI Appendix, Table S2 and illustrated in Fig. 2. We take these calculated liquid concentrations as approximations of the C contents of the core portions of the parent planetesimals (See SI Appendix Text for further discussion). In the *log* C/S versus *log* C diagram, calculated parent cores fall chiefly into two different fields based on C/S ratio (Fig. 3), which we term “medium C/S” (0.008 < C/S < 0.03) and “low C/S” (C/S < 0.006), plus two higher C/S outliers. Note that the same meteorite groups plot in different locations in this plot, depending on the source of estimated C and particularly S (SI Appendix, Table S2), the latter of which varies considerably depending on whether the methodology of Chabot or Wasson was employed. For example, the IVA irons are located in markedly different locations on the plot, depending on the S concentration estimate employed (Fig. 3, SI Appendix, Table S2). However, nearly all such estimates result in core compositions with medium C/S or low C/S. The two high C/S outliers are considered further in the Discussion.

**Planetesimal volatile loss processes**

### ***Processing on planetesimals similar to chondrites***

Heating of primitive bodies is accompanied by aqueous fluid transport and volatile loss (60, 61). The resulting effects on C and S can be gauged empirically from observed *log* C/S versus *log* C variations with petrologic type (Fig. 1B, SI Appendix, Table S1). Significant depletions in C with petrologic type are unsurprising, given that metamorphism, differentiation, and devolatilization, possibly including interior melting (60), occurred on chondrite parent bodies (62). S losses are evident for samples of higher petrologic type (e.g., the CK group). These metamorphic effects are consistent with observed decreases in highly volatile elements in ordinary chondrites with increasing petrologic type, without appreciable changes in moderately volatile concentrations (62).

The depletion of volatiles in chondritic bodies with increased textural equilibration must also have occurred during the early heating of planetesimals prior to separation of metal-rich melts that produced differentiated mantles and cores. This means that the immediate precursors to iron parent planetesimals were likely already partially devolatilized. The extent of devolatilization within a maturing planetesimal is not well understood, as it depends on location within bodies of varying radius, as well as the coevolution of thermal state, stresses, mineral reactions, and permeability (60, 61). In the following models of planetesimal differentiation, we consider three hypothetical precursor materials: a volatile-rich chondrite (VRC) similar to CI and CM groups, a volatile-depleted chondrite (VDC) similar in C and S concentrations to chondrites with petrologic types 5 or 6 or to acapulcoites, and partially-depleted chondrite (PDC), intermediate between the VRC and VDC and somewhat more enriched than enstatite chondrites or chondrites of intermediate petrologic type (H3, CV3, etc.) (Fig. 1).

***Processing on differentiated planetesimals***

To explore the effects of planetesimal differentiation on C/S-C systematics of putative planetesimal cores, we model different scenarios, corresponding to progressively greater heating (Fig. 4). The simplest (Fig. 4A1) is segregation of a metallic core in a closed-system by complete melting of the alloy but without appreciable melting of the silicate. A second case (Fig. 4A2) is formation of a planetesimal magma ocean beneath a solid impermeable outer shell (63, 64) in which molten alloy and silicate equilibrate without degassing to the surface. The third (Fig. 4B) and fourth (Fig 4C) cases include formation of silicate melt in processes that allow surface degassing, with subsequent loss of this atmosphere. In the third case (4B), degassing is limited and could be caused by volcanic eruptions or by impacts. Impacts can incite degassing by formation of limited surficial magma ponds (65) or by post-impact excavation of the interior (66, 67). Partial degassing is also expected in the multi-stage planetesimal core formation models proposed by Neumann et al. (68). In the fourth case (4C), degassing of the silicate portion of the planetesimal occurs owing to formation of an uncovered magma ocean (13). Formation of a wholly molten planetesimal is favored for large (>300 km) bodies formed within the first 0.5 Ma of solar system history owing to heating from $^{26}Al$ (13, 60). Fractionation of Mg and Si isotopes between silicate achondrite (HED, angrite) parent bodies relative to chondrites (24, 26) apparently requires significant high temperature degassing consistent with extensive near-surface magma (13).

For the case in which only alloy melts, C and S in the resulting core are controlled by the parent planetesimal bulk composition and the relative masses of silicate and metal. This assumes that no phases capable of storing appreciable C and S are retained in the silicate, but as discussed below and in the SI Appendix Text, some C may be retained in the silicate shell as graphite. Removal of all C and S to the core produces a liquid with the same C/S ratio as the bulk planetesimal, with C enhanced according to the inverse of the metal fraction in the planetesimal (Fig. 4A). For example, cores comprising 5-30% of the planetesimal mass with the PDC composition have 16-2.7 wt.% C.

Models involving silicate melting require assumptions about relevant metal/silicate partition coefficients, which depend on oxygen fugacity and are detailed in Table S3, and the degree of metal/silicate equilibration (69). In most scenarios, C is more siderophile than S (34), though in the case of segregation of S-rich (> 18 wt.% S) cores (applicable to VRC and PDC, but not VDC), the opposite relation holds (70) (Table S3).

Cores from a sealed magma ocean are enriched in C and have C/S ratios similar to or greater than bulk planetesimal compositions, unless the cores are highly enriched in S, in which case the C/S ratios are lower than their source (Fig. 4A). Cores derived from planetesimals which have partially degassed during silicate melting are commensurately less enriched in C, with the specific C/S ratios controlled partly by the magnitude of C versus S outgassing (Fig. 4B). Cores formed from planetesimals that undergo wholesale melting have the lowest total C contents and also have markedly reduced C/S, owing to the much greater solubility of S in silicate melts relative to C (Fig. 4C).

Some of the calculated core compositions in Fig. 4, and particularly those derived from closed-system differentiation, plot within the fields of stability of 2 liquids or of graphite. Such liquids would partly crystallize or unmix, producing liquids at the boundaries of the graphite and 2-liquid stability fields, respectively. Separation of S-rich liquids from graphite-saturated planetesimal mantles could leave behind C-enriched silicate residues with some similarities to ureilite achondrites (See *SI Appendix Text* for further discussion).

## Discussion

### *Parent body cores produced from degassed of devolatilized planetesimals*

Modeled closed system differentiation of planetesimals produces cores that are more enriched in C than those inferred than the medium and low C/S parent body cores (Figs 4A and 4B). For the less-degassed VRC and PDC compositions, modeled cores have >20 times the C concentrations found in groups with medium and low C/S. For the much less C-rich VDC composition, modeled closed-system core compositions remain at least twice as C-rich. Prior to or coeval with core formation, planetesimals originating from material similar to chondrites and now represented by iron meteorites must have experienced considerable loss of highly volatile elements beyond that represented by even volatile-poor chondrites.

As noted above, model core compositions resulting from closed-system differentiation largely have compositions that are expected to unmix to C-rich and S-rich liquids, particular in the case of less degassed (VRC and PDC) bulk planetesimal compositions (Fig. 4A). For these, the S rich conjugate liquids have >0.1 wt.% C, and so are enriched in C compared to the parent body cores with medium C/S by factors of 2-10 (Fig. S2). Origin of the medium C/S parent body cores simply by unmixing therefore seems unlikely. Once segregated to a core, a mechanism for secondary volatile loss would be required, but as argued below this additional process would have to be mediated by silicates, as devolatilization of bare iron cores would deplete S with little change in C and could not account for the observed C deficit.

The parent body cores with medium C/S can be reproduced if considerable silicate melting and degassing to the surface occurred prior to or during core formation, but only if the initial planetesimal had already been largely degassed, similar to the VDC composition (Fig. 4B). However, the low C/S parent bodies are so C-depleted that they require even more extensive degassing processes, as may have occurred in whole-planetesimal magma ocean scenarios (Fig. 4C).

***Importance of degassing of silicates***

A key inference is that evident planetesimal core volatile depletion occurred chiefly by loss from degassing of silicate or, prior to core formation, from silicate-metal portions of planetesimals. The metamorphic loss of volatiles that accounts for the chondrite trend (Fig. 1B) is owing to decomposition of accessory phases (mainly organics, sulfides) in a silicate-alloy matrix (62, 71, 72). The more advanced volatile loss and diminished C/S evident in the low C/S parent body cores is a hallmark of molten silicate degassing, owing to the greater solubility of S compared to C in such melts. Degassing of iron cores in the absence of silicate would have the opposite effect, as the vapor pressure of S above molten Fe-C-S alloy is orders of magnitude greater than that for C (Fig. 4C and see *SI Appendix Text*). Although some loss of volatiles from cores in the absence of their silicate mantles is not precluded, and has been inferred based on Pd-Ag isotopes for the IVA group (73), it would result in extensive loss of S without appreciable loss of C (Fig. 4C and *SI Appendix Text*) and so cannot be the explanation for the low C/S of parent cores. We conclude that the parent planetesimals that produced extant iron meteorite groups were strongly depleted in carbon.

***Effect of conflicting estimates of core S content and significance of high C/S outliers***

The approach adopted here is to explore diverse and conflicting estimates of S and C for iron meteorites and for their planetesimals, and we find that the resulting inferences and conclusions do not depend on which sets of values are preferred. Irrespective of whether S-poor or S-rich parental metallic liquids for a particular iron group are accepted, nearly all calculated planetesimal cores fall either into the medium C/S or low C/S fields and both are populated by estimates that come from each methodology (Fig. 3). Thus, the conclusion that planetesimals parental to iron cores experienced significant degassing does not depend on establishing which methodology is more accurate.

Two apparent exceptions are high C/S outliers ($IVA_W$ and $IID_W$ in Fig. 3 and SI Appendix, Table S2) for which S contents of 0.4 and 0.7 wt.% were calculated, respectively (45, 46) and which cannot be solely the products of the degassing processes modeled in Fig. 4. One interpretation is that the S concentrations of these are underestimated, perhaps because the melt/liquid partition coefficient of Ir, a key element in constraining liquid S content, was overestimated as compared to experimental results (39), resulting in underestimates of liquid S.

A second interpretation, offered to account for an inferred S-poor core for the otherwise volatile (Ge, Ga)-rich IID body, is two-stage disequilibrium core formation, resulting in a S-rich outer core and S-poor inner core (46). However, this process alone could not account for the C concentrations of the IID body. Significant parent body outgassing of C prior to core segregation would also be required. Otherwise the resulting S-poor inner core would be C-enriched, meaning it would plot to the C-rich side of the "chondrite trend" on Fig. 3.

A third possibility is outgassing of an exposed core from a planetesimal that had lost its silicate shell. As described in the previous section, this, would deplete the core in S with negligible S loss, and thereby raise C/S and such a process could be relevant to the IVA group based on Pd-Ag isotope systematics (73).

***Significance for volatile delivery to planets***

Compared to precursors similar to chondrites, inferred planetesimal core compositions indicate that planetesimal differentiation was associated with significant volatile loss. The isotopic kinship between iron meteorites and chondrites, and in particular observation that some of the metallic cores that are highly depleted in C (e.g., groups IIC and IVB) have carbonaceous chondrite (CC) parentage (32), makes clear that this depletion occurred on planetesimals, and was in addition to the devolatilization that occurs in primitive materials prior to planetesimal accretion (6). As documented in the modeling above, the volatile depletion processes, particularly those that resulted in low C/S, are characteristic of silicate degassing, which means that the inferred depletion affected the mantles of these planetesimals as well as their cores. The processes responsible for depletion of C, including metamorphic destruction of carbonaceous carriers in undifferentiated parent bodies and magmatic degassing into tenuous atmospheres, should also have pronounced effects on other highly volatile elements including nitrogen and hydrogen.

Isotopic evidence (33) and dynamical modeling (74) indicate that the parent body cores segregated from planetesimals and embryos born during early and efficient aggregation, whereas materials now preserved as chondrites formed late from remnant materials, possibly accreting as crustal veneers to earlier-formed differentiated planetesimals (33). Early formation of iron meteorite parent bodies and Mars (75), along with astronomical evidence for rapid drops in the observable mass of solids in disks (76) suggest that a large fraction of the raw materials for accreting planets was incorporated into planetesimals and embryos on ~100 kyr timescales. Owing to $^{26}Al$, such bodies would have reached high temperature, and thus undergone the processing described here. This effect should be more pronounced near 1 AU than in the asteroid belt, owing to more rapid planetesimal accretion time scales (77, 78). Therefore, if embryos and planets accreted chiefly from planetesimals, rather than from pebbles, then the largely devolatilized planetesimals evidenced from parent body cores are likely better models for planetary feedstocks than chondrites. Lambrechts et al. (79) suggested that the flux of pebbles in the inner disk might lead to two modes of planet formation. Super-earth dominated systems form in the case of high pebble flux, whereas the low pebble flux regime favors formation of Mars-sized embryos, which in turn lead to giant-impact creation of terrestrial planets. The lack of super-earths in our Solar System suggests that this latter solution, with its greater role for planetesimals compared to pebbles, may be most relevant for the effects described here.

These considerations highlight that a significant fraction of the terrestrial planets were likely derived from differentiated objects that were unlike chondrites in their volatile cargos. This was particularly so for objects accreted during the early stages of planet formation and perhaps less prevalent for later-added materials arriving from high heliocentric distance (80, 81), though evidence from parent body cores shows that early-formed planetesimals derived from carbonaceous chondrites also were degassed. This may have limited the supply of S and particularly C to cores of these planets. More generally chondrites are probably poor guides for the sources of planetary volatiles and the common modeling of planetesimals and embryos with chondritic compositions to account for the accretion of major volatiles to terrestrial planets (e.g., 70, 80, 81) is questionable. Recently, Piani et al. (82) suggested that about 3 oceans of $H_2O$ could have been delivered by material similar to enstatite chondrites throughout its accretion, but this does not account for loss of $H_2O$ during planetesimal or embryo differentiation, and so is likely an overestimate.

**Conclusions**

Loss from planetesimals is an important stage in the evolution from the volatile-rich solar nebula to volatile-poor, potentially habitable rocky planets. Iron meteorites provide a useful record of volatile processing and loss during planetesimal differentiation. Carbon concentrations in planetesimal cores represented by magmatic iron meteorite parent bodies can be estimated from the compositions of iron meteorites by assuming the latter are cumulates formed from the parent liquid and applying thermodynamic models for solid and liquid Fe-alloys and Fe-S-C liquid. Though the parent bodies are known to be variably enriched in S, they are *all* poor in C. This necessitates volatile carbon loss in the inner solar system via planetesimal and nebular processes.

Modeled closed system core formation from chondrite-like planetesimals are significantly more enriched in C than inferred parent bodies of iron meteorites, highlighting that planetesimals experience open-system outgassing. Devolatilization comparable to that evident from high petrologic type chondrites is insufficient to account for the magnitude of carbon loss evident from the irons, indicating that silicate melting played a role in planetesimal evolution, either by surface magma oceans or more limited devolatilization processes, perhaps associated with impacts. Substantial C-depletion evident in parent body core compositions, but with significant remaining S, is consistent with volatile loss controlled by devolatilization and melting of silicates, but not with degassing of bare iron cores. Thus, the volatile-depleted character of parent body cores reflect processes that affected whole planetesimals. As the parent bodies of iron meteorites formed early in solar system history and likely represent survivors of a planetesimal population that was mostly consumed during planet formation, they are potentially good analogs for the compositions of planetesimals and embryos accreted to terrestrial planets. Less-depleted chondritic bodies, which formed later and did not experience such significant devolatilization, are possibly less apt models for the building blocks of terrestrial planets. More globally, the process of terrestrial planet formation appears to be dominated by volatile carbon loss at all stages, making the journey of carbon-dominated interstellar precursors (C/Si > 1) to carbon-poor worlds inevitable.

## Acknowledgements

We are grateful for the detailed and helpful reviews from Nancy Chabot and Rich Walker. This research comes from an interdisciplinary collaboration funded by the National Science Foundation INSPIRE program, through grant AST1344133. Additional funding has been provided by NASA 80NSSC19K0959 (to MMH), XRP, NNX16AB48G (to GAB), and XRP, 80NSSC20K0259, to EAB and FJC.

**Figures**

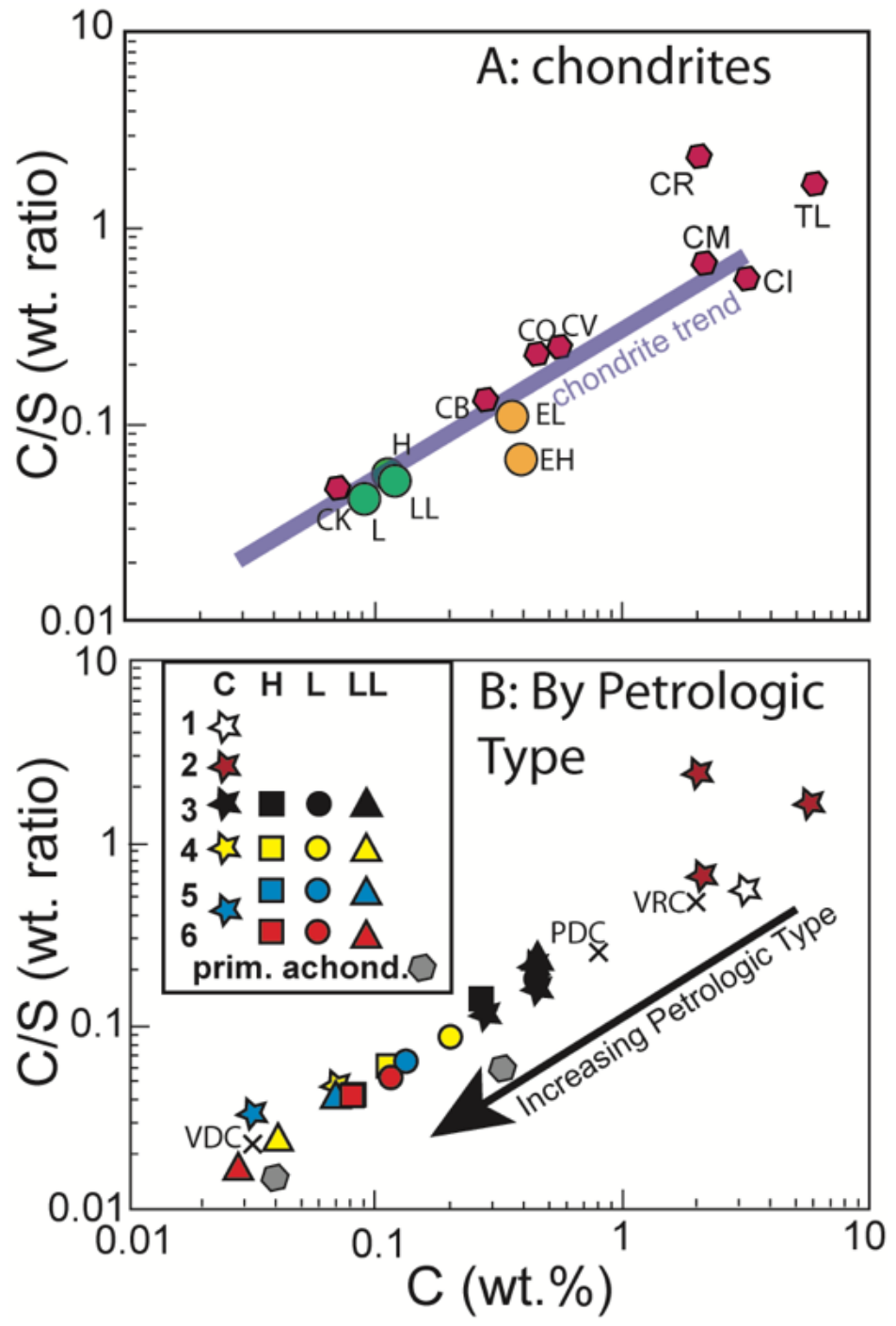

**Fig. 1** C and S compositions of chondrite meteorite groups plotted as *log* C/S versus *log* C, following Hirschmann (34). (A) average compositions of different chondrite groups (B) average compositions of carbonaceous and ordinary chondrites, sorted by petrologic type, plus average compositions of primitive achondrite groups, acapulcoites and winonaites. Data sources are given in SI Appendix, Table S1. Also shown are 3 model compositions used in calculations: VRC (volatile-rich chondrite), PDC (partially degassed chondrite), and VDC (volatile-depleted chondrite).

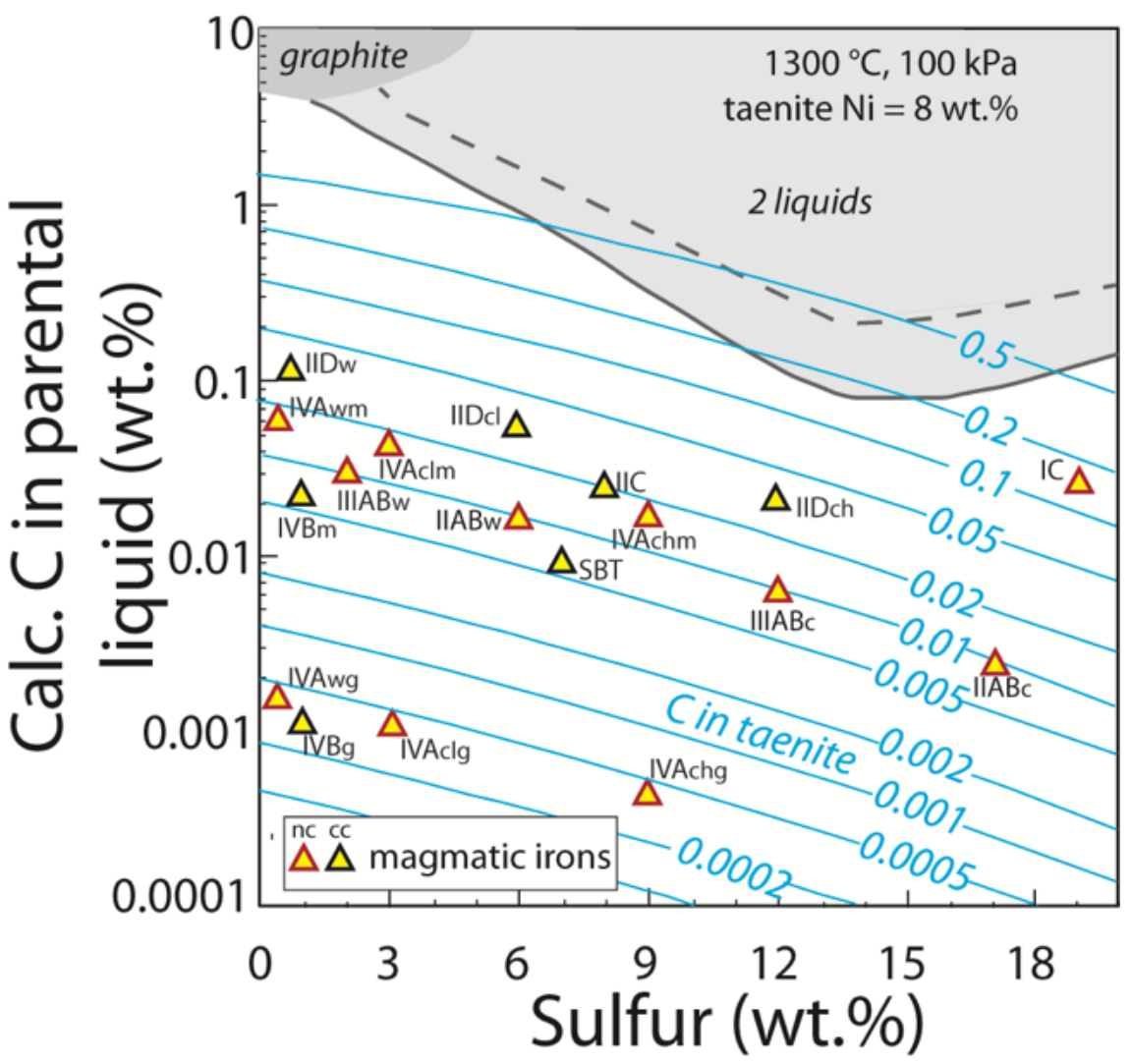

**Fig. 2** C concentrations of planetesimal cores calculated from magmatic iron meteorite groups, using their inferred sulfur contents (SI Appendix, Table S2) and assuming that the C contents of iron meteorites (also SI Appendix, Table S2) represent pure cumulate taenite compositions (blue curves). Subscripts (key in SI Appendix, Table S2) denote different published estimates of S and/or C, as described in main and SI Appendix Text. From the assumed taenite composition, the activity of C is calculated from the thermodynamic model for Fe-Ni-C (50) at 100 kPa, 1300°C, assuming 8 wt.% Ni. Higher temperatures would make C more compatible in the solid, resulting in lower calculated liquid C and the calculations are not strongly dependent on the assumed Ni content for ±5 wt.% (See Fig. S1). From this C activity, the liquid C concentration is calculated in Fe-C-S liquid with the specified S concentration from the model of Wang et al. (53). The fields of immiscible liquids and graphite saturation in the system Fe-C-S are shown for the 100 kPa liquidus surface from the thermodynamic model of Tafwidly and Kang (83). The dashed line is the estimated limit of the 2-liquid field for 10 wt.% Ni, based on experimental data (56, 84). No inferred core compositions are consistent with equilibrium with a second alloy liquid or graphite.

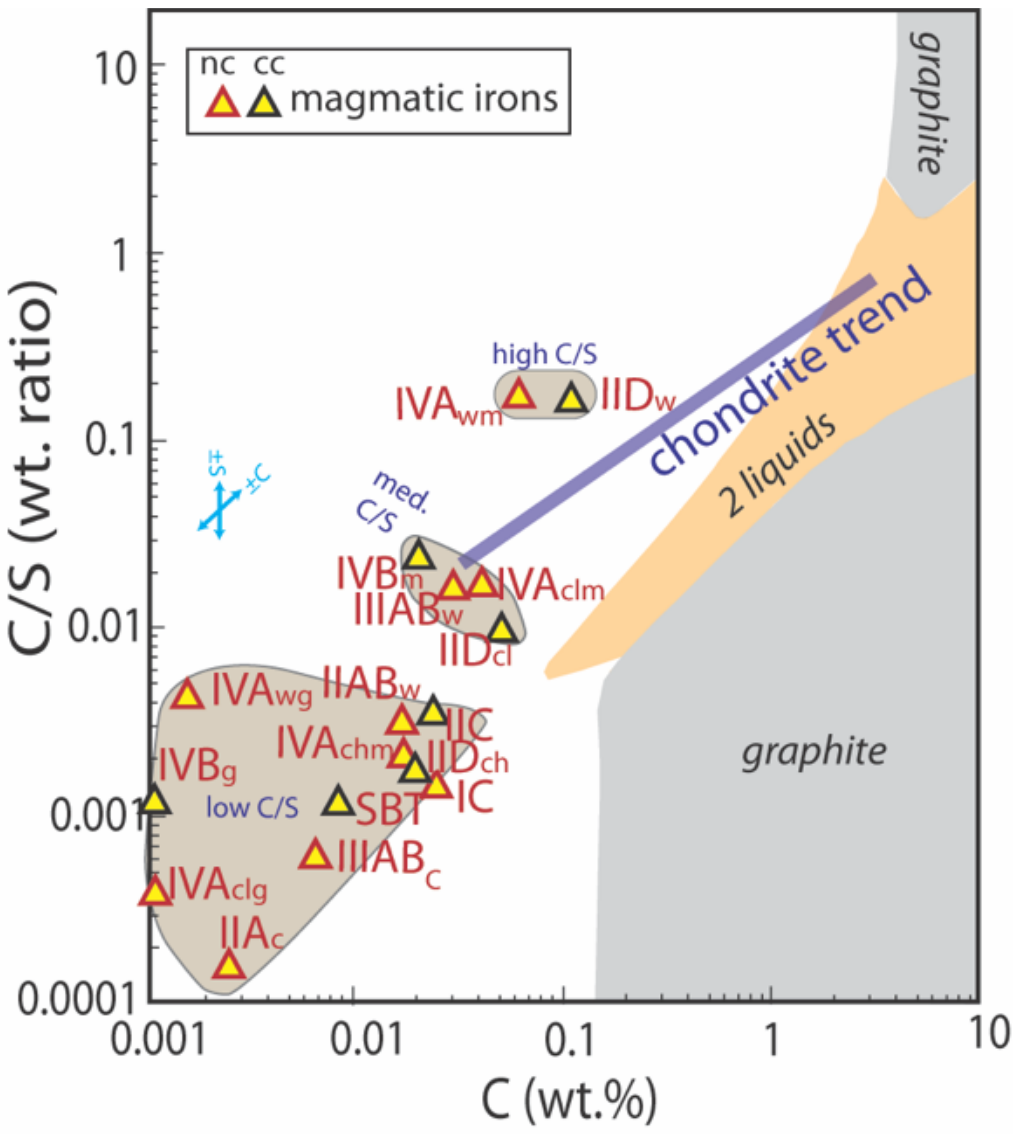

**Fig. 3** C/S versus C compositions of planetesimal cores calculated for magmatic iron meteorite groups (SI Appendix, Table S2, Fig. 2). Subscripts (key in SI Appendix, Table S2) are as described in caption to Fig. 2. The diagonal line shows the trend of chondritic compositions, from Fig. 1. Fields of immiscible liquids and graphite saturation in the system Fe-C-S are shown for the 100 kPa liquidus surface from the thermodynamic model of Tafwidly and Kang (83). Small blue arrows illustrate the effect of independent variation of C or S concentrations on the C/S versus C plot.

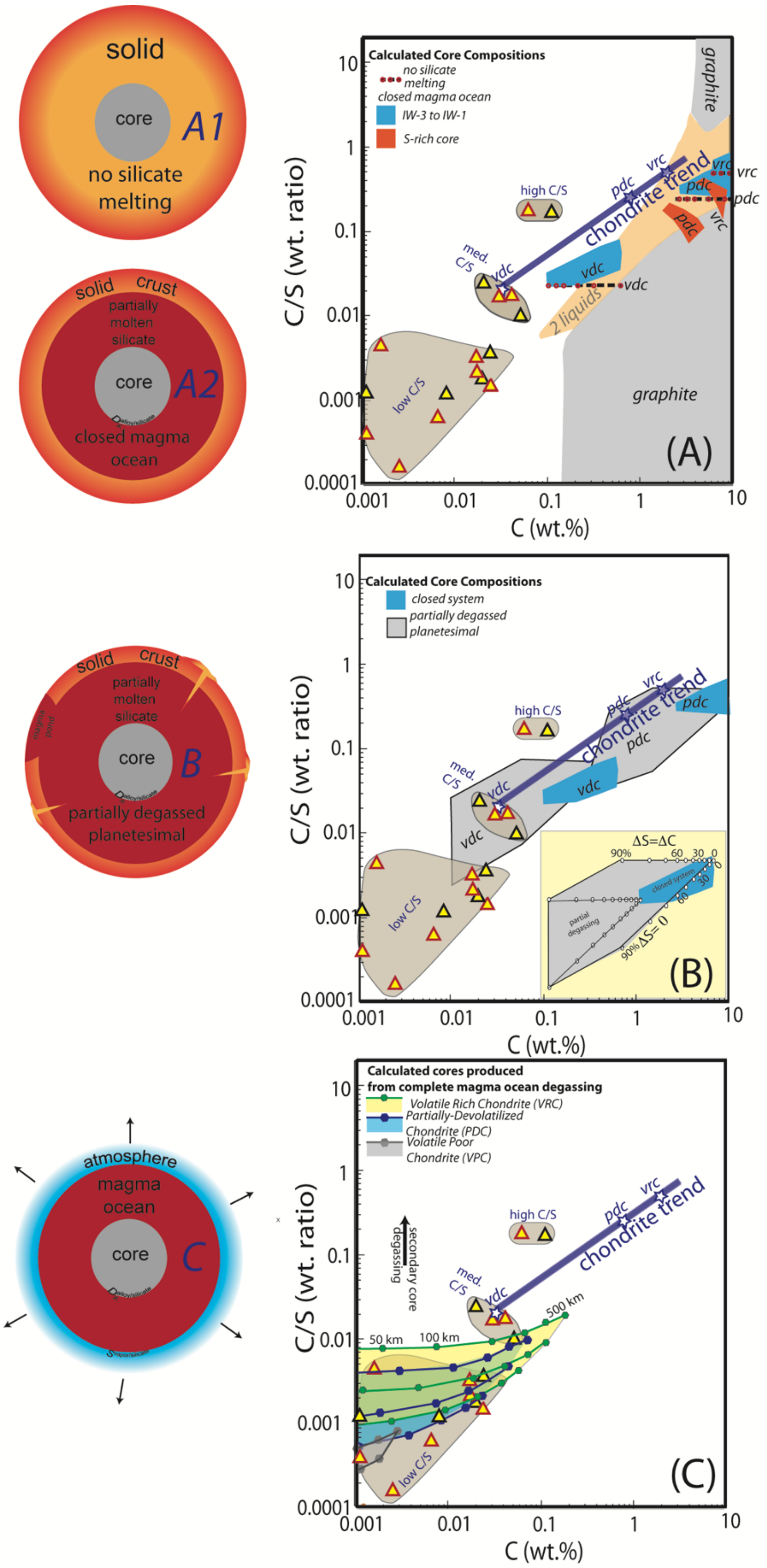

**Fig. 4 Caption on following page**

**Fig. 4** Modeled core formation effects on planetesimals from chondritic precursors compared to inferred compositions of parent body cores (SI Appendix, Table S2, Fig. 3). Three chondritic precursor compositions, VRC (volatile-rich chondrite), PDC (partially depleted chondrite), and VDC (volatile depleted chondrite) are used (SI Appendix, Table S1, Fig. 1). For labels to individual iron meteorite groups, see Fig. 3. In (A), two different core-forming processes are modeled. "No silicate melting" [cartoon *A1*] shows removal of alloy+C+S from each bulk composition planetesimals ranging from 5% (more C-rich cores) to 30% metal (less C-rich). "Closed magma ocean" [cartoon *A2*] shows removal of core alloy from a planetesimal with a solid carapace and interior magma ocean (63), with no degassing to the surface. Calculated core compositions are for planetesimals with 5-30% metal, partition coefficients given in Table S3, and extent of metal-silicate equilibration ("*Q*" in Eqn. S1) ranging from 50-100%. Partition coefficients are appropriate for core formation at oxygen fugacities of IW-3 to IW-1, as well as for formation of S-rich (>15 wt.%) cores. The latter are not calculated for the VDC composition, which is too S-poor to generate S-rich cores. Some of the calculated core compositions plot within the stability fields of graphite or of 2 liquids (see SI Appendix Text). Part B models a "Partially degassed magma ocean" (calculated only for PDC and VDC compositions) which is similar to the "Closed magma ocean" calculations except that the molten silicate is assumed to have partially degassed to the planetesimal surface [cartoon *B*]. Inset illustrates this, which is assumed to be loss of 0-90% of either the C ("$\Delta$=0", diagonal bound on each trend) or both the C and S ("$\Delta S$=$\Delta C$", horizontal bound), as further described in the SI Appendix Text. (C) Modeled cores produced by segregation from a planetesimal in which alloy was segregated from a magma ocean that had equilibrated with an atmosphere produced by whole-planetesimal degassing [cartoon *C*]. Larger planetesimals produce atmospheres with greater partial pressures, thereby enhancing volatile retention in silicate and alloy portions. For each planetesimal bulk composition (VRC, PDC, VDC), calculations are conducted for planetesimals with diameters of 500, 400, 300, 200, 100, 50, 20, and 10 km (with symbols in plot moving from C-rich to more C-poor as diameters diminish) and for partition coefficients appropriate for IW-3 (highest C/S for each bulk composition), IW-2, and IW-1 (lowest C/S). The vertical arrow shows the putative effect of secondary degassing of bare iron cores after differentiation and fragmentation of the parent meteorite. This results in extensive loss of S and negligible C loss, increasing the C/S ratio at constant C concentration.

# Supplementary Information for

## Early volatile depletion on planetesimals inferred from C-S systematics of iron meteorite parent bodies

**Marc M. Hirschmann[1,*], Edwin A. Bergin[2], Geoff A. Blake[3], Fred J. Ciesla[4], Jie Li[5]**

[1] Department of Earth and Environmental Sciences, University of Minnesota, Minneapolis, MN 55455.
[2] Department of Astronomy, University of Michigan, Ann Arbor, MI 48109.
[3] Division of Geological & Planetary Sciences, California Institute of Technology, Pasadena, CA 91125.
[4] Department of Geophysical Sciences and Chicago Center for Cosmochemistry, University of Chicago, Chicago, IL 60637.
[5] Department of Earth and Environmental Sciences, University of Michigan, Ann Arbor, MI 48109.
*Corresponding author: M.M. Hirschmann. mmh@umn.edu

**This PDF file includes:**

Supplementary text
Figures S1 to S2
Tables S1 to S3

**Other supplementary materials for this manuscript include the following:**

Excel spreadsheet for the thermodynamic model used to calculate the C content of Fe-C-S alloy liquids from coexisting Fe-Ni-C solid alloy in Table S2 and Figures 2 and S1. File Name: HirschmannPNAS-CFig2+S1Calc.xlsx

**Carbon in iron meteorite groups.**

We employ C estimates for 8 magmatic iron meteorite groups (Table S2). Carbon in iron meteorites has been evaluated by bulk analyses (1, 2), by point counting or planimetric analysis of polished slabs (3), and by combining microanalysis with modal analysis (4). In one case, Meibom et al. (5) determined bulk C of a suite of IIIAB irons using nuclear reaction analysis. Groups not included in our compilation were omitted simply because either sulfur or carbon estimates are lacking.

In contrast to sulfur, which is hosted chiefly in accessory phases, carbon in iron meteorites is hosted in FeNi alloy as well as graphite and carbide (4). Within magmatic groups, C concentrations do not show evidence of systematic increases or decreases with indices of differentiation, such as Ni content (4). Consequently, C concentrations indicative of a magmatic group can be inferred from a sampling of individual iron meteorites, without consideration of their placement in a fractional crystallization sequence.

The most comprehensive survey of carbon in iron meteorites comes from the 130 bulk compositions determined by combustion analysis (1, 2). These span the range from 0 to 0.45 wt.% C. Unfortunately, the analyses were specifically targeted to portray the composition of meteoritic metal, and were conducted by bulk analysis of sections of material from which visible inclusions were excluded. As described by Moore et al. (2), this approach eliminated most large grains of graphite and carbide evident to the naked eye, but not finer grains occurring interstitially or as inclusions. Consequently, these analyses underestimate C in irons and iron groups with conspicuous C-rich domains, but can be more accurate for those in which accessory carbon minerals are absent or limited to microscopic interstitial phases.

Buchwald (3) provided bulk estimates of carbon in selected irons based on planimetric analyses and point counting of large etched slabs. Perhaps even more valuably, Buchwald (3) also gave detailed petrographic descriptions of hundreds of irons that were then known. These reveal which irons and groups have large segregations of graphite or carbide evident without the aid of microscopy, have C phases observable only on microscopic investigation, or have no discernable C rich accessory phases at all.

Goldstein et al. (4) presented the most comprehensive modern survey of C in iron meteorites, combining SIMS and EPMA analyses and imaging with modal analyses to construct characteristic C estimates for 4 of the 8 iron groups (IIAB, IIIAB, IVA, IVB) included in the present study. The following paragraphs provide a more detailed discussion of these four groups, comparing Goldstein et al.'s estimates with other available constraints. We also describe evidence for C contents of the groups (IC, IIC, IID, SBT) not investigated by Goldstein et al. (4).

**IC** We estimate the C concentration for the IC group of 0.15 wt.% from the average of point-counted estimates by Buchwald (3) of meteorites Bendego (0.1 wt.%), Chihuaha City (0.15 wt.%), and Santa Rosa (0.19 wt.%).

**IIC** We estimate the C concentration of the IIC group of 0.019 wt.%, based on the analyses of Bacrubito, Perryville, and Wiley (4). Detailed petrographic examination of these 3 meteorites by Buchwald (3) did not detect any graphite or carbides, suggesting that the combustion analyses are not biased by selective exclusion of C-rich phases, and that the low average concentration is realistic.

**IIAB** Goldstein et al. (4) estimated a C content of the IIAB group of 0.01 wt.% based on SIMS analysis of a single meteorite, North Chile, and for this sample, most of the C was considered to reside in small (1 mm) graphite inclusions observed petrographically by Buchwald (3). Bulk analyses of 19 IIAB irons (4) are consistent with this estimate (0.008±0.004 wt.%). Roughly half of the petrograpic analyses of IIABs reveal small 10-100 micron segregations of cohenite±graphite, at times attached to mm-sized sulfide grains (3), whereas no petrographically observed C-rich phases are observed in the others. The latter include the intensively examined Sikhote Alin and also Silver Bell, the IIAB for which Moore et al. (2) and Lewis and Moore (1) report their highest C content (0.22 wt.%). It is difficult to say whether the bulk chemical analyses (1, 2) were on samples that selectively avoided these inclusions, but the consistency of their analytical results suggests that this was not a major factor and that the estimate of Goldstein et al. (4) is reasonable.

**IID** The average C concentration of 3 IIDs (Carbo, Rodeo, Wallapai) analyzed by (1, 2) is 0.03 wt.% which is consistent with the petrographic estimate (also 0.03 wt.%) of IID N'Kandhla by Buchwald (3). Buchwald's petrographic examination of several IIDs (Bridgewater, Brownfield, Needles, Losttown, Mount Ouray, Puquios, Richa, Wallapai) did not reveal any C-rich phases, but microscopic needles of graphite or carbide rosettes were identified in Carbo, Elbogen, and N'Kandhla. We infer that the combustion analyses are not biased by selective exclusion of C-rich phases, and that the average concentration is realistic.

**IIIAB.** Goldstein et al. (4) estimated average C concentrations for the IIIAB group to be between 0.001 and 0.01 wt.%. The upper end of this range is consistent with combustion analyses of 52 IIIAB irons that average 0.016±0.012 wt.% (1, 2) and with the average of nuclear reaction analyses of 7 IIIAB irons of 0.0065±0.0035 wt.% (5). Significant C-rich heterogeneities are not likely a problem for this group. For example, no C-rich phases have been noted in extensive studies of very large polished slabs of Cape York (3). We therefore take 0.01 wt.% as a reasonable estimate for the IIIAB group.

**IVA and IVB** The IVA and IVB groups are strongly depleted in moderately volatile elements Ge and Ga (6) and clearly have very low total carbon, yet quantitative estimates differ significantly. Goldstein et al. (4) report 0.0004 wt.% and 0.0003 wt.% C in groups IVA and IVB, respectively, based on examination of one meteorite (Bishop Canyon, Tawallah Valley) from each group. In contrast, average analyses from (1, 2) for IVA and IVB groups are 0.016 and 0.006 wt.% C from 15 and 4 meteorites, respectively. Goldstein et al. (4) suggest that these larger averages are a consequence of small amounts of contamination in the combustion analyses. In the case of the IVA group, C concentrations for the 15 meteorites range from 0.003 to 0.042 wt.% (1, 2), and it is the more C rich meteorites that are both most responsible for the relatively high average C concentration and least likely to be impacted by contamination. The four IVB irons range from 0.004 to 0.11 wt.% C (1, 2). Also, for the 19 IVA and IVB irons analyzed by combustion, analyses on duplicate aliquots were performed and are reported individually for the 16 that come from the study of Moore et al. (2). Where reported, the duplicates closely reproduced one another. Though it is quite probable that the combustion analyses were inflated slightly from analytical blanks comparable to the lowest reported concentrations for individual analyses (0.003-0.005 wt.% C), these would not be nearly sufficient to account for the differences with the SIMS study of Goldstein et al. (4), at least for the IVA group. Three of the four irons from the IVB group are close to this minimum threshold, but the fourth, Deep Spring, with 0.011 wt.%, is greater. At the same time, the average concentrations derived by Goldstein et al. (4) for groups IVA and IVB come from just 6 analyzed SIMS spots on Bishop Canyon and 4 on Tawallah Valley, and therefore it is possible that local microscopic concentrations of C, averaged into the bulk combustion analyses, were missed by the microanalytical approach. On balance, we believe that the average combustion analyses may be more characteristic of the IVA and possibly the IVB groups, though we also consider the more extremely depleted alternative values from Goldstein et al. (4).

**SBT** The South Byron Trio is a small grouplet of irons (7, 8). We estimate a C concentration of 0.006 wt.%, based on combustion analysis of South Byron (2). Low estimated C concentrations are consistent with petrographic examination, as both conventional metallography (2), and detailed electron microscopy (8) have not revealed any macroscopic or microscopic C-rich phases.

### Partial Degassing of Planetesimal Interiors

Partial degassing of planetesimal interiors without wholesale melting and formation of surface magma oceans is the third differentiation process discussed in the main text and is illustrated in Fig. 4B. It could occur either owing to impacts or to igneous crust formation. Of these, degassing associated with impacts is likely more effective, as it promotes (localized) heating that can induce metamorphic devolatilization over a wide area or high degree partial melting in a more limited domain. In the latter, volatile loss is controlled by magmatic degassing. In contrast, volatile dep;etion from planetesimal interiors by extraction of silicate melt is controlled by partitioning between residual mantle phases and silicate magma. This is comparatively inefficient. For example, under even modestly reduced conditions, the concentration of C in graphite-saturated magmas is severely limited (9, 10), impeding significant C extraction. Solubility of S under reduced conditions can be much greater, up to ~1 wt. % (11), but this will only diminish the S concentration of the residual interior if the initial concentration is less than that of the magma. In such a scenario, S would behave as a mildly incompatible element. But even highly devolatilized chondrites have >1 wt.% S (e.g., 1.5 wt. % S in VDC, Table S1), meaning extraction of a melt with 1 wt.% S would leave behind a mantle with *increased* bulk S concentration.

The partial degassing trends depicted in Fig 4B span a range of behavior from loss of C and full retention of S ($\Delta$S=0, inset to Fig. 4B) to equal loss of S and C ($\Delta$S=

$\Delta$C). These limits were chosen owing to the greater solubility of S relative to C in silicate magmas (Table S3), meaning that degassing should affect C more than S. In the case of igneous extraction of melt from a solid matrix under reducing conditions, the opposite relationship can hold, with retention of C and loss of S. This could lead to increases in C/S with modest changes in C. However, under reduced conditions, the efficacy of this mechanism is limited unless the source is already highly depleted in S. For example, extraction of 20% basalt with negligible C and 1 wt.% S from a VDC planetesimal interior (1.5 wt.% S, 0.022 C/S; Table S1) would produce a residuum (1.6 wt. % S; 0.026 C/S) that was little-changed. Extraction of the same basalt from a much more depleted source could have a larger effect, e.g., if the initial source had 0.01 wt.% C and 0.3 wt.% S, the C/S ratio would go from 0.033 to 0.1.

## Vapor loss from iron cores without silicate mantles

One process suggested to account for volatile depletion of some parent body cores is re-melting of preexisting metallic cores that had previously lost their silicate mantles. Pd-Ag isotopes provide evidence for such a process for group IVA (12, 13), which is highly depleted in both C and S (Fig 4; Table S2). However, owing to the large disparity in vapor pressures between S and C above molten iron, it is S, rather than C that would be lost from such a process. For example, at 2000 K, an Fe-S melt with mole fraction of S equal to 0.1 has a S vapor pressure of ~1000 Pa (14), whereas the C vapor pressure of a graphite-saturated iron alloy melt would be eight orders of magnitude lower ($10^{-5}$ Pa; (15)), and an unsaturated melt would have C vapor pressure commensurately lower (with vapor pressures of both elements calculated as monatomic species). Consequently, evaporation of a hot iron mass into a vacuum would deplete residual iron bodies in S and have little effect on their C contents, causing increased C/S at near-constant C (Fig. 4C).

## Calculated C partitioning between taenite and Fe-C-S melts

Calculations of partitioning of C between taenite and coexisting alloy melt (Fig. S1) are performed by fixing the solid alloy composition and calculating the activity of C using the model of (16) and then at fixed wt.% S in the liquid, searching iteratively for the C concentration in liquid Fe-C-S that matches that activity, using the model of (17). For both models, C activities are relative to a standard state of unity for graphite saturation. Increased S in the liquid greatly enhances the activity coefficient of C in the liquid, thereby stabilizing C in the solid relative to the liquid. The role of Ni in the liquid is ignored, but because Ni diminishes the concentration of C required to saturate graphite by about 20% (relative) for 20 wt.% Ni (18, 19), its effect would be to increase slightly calculated values of $D_C$ and therefore to decrease the inferred C contents of iron parent body liquids.

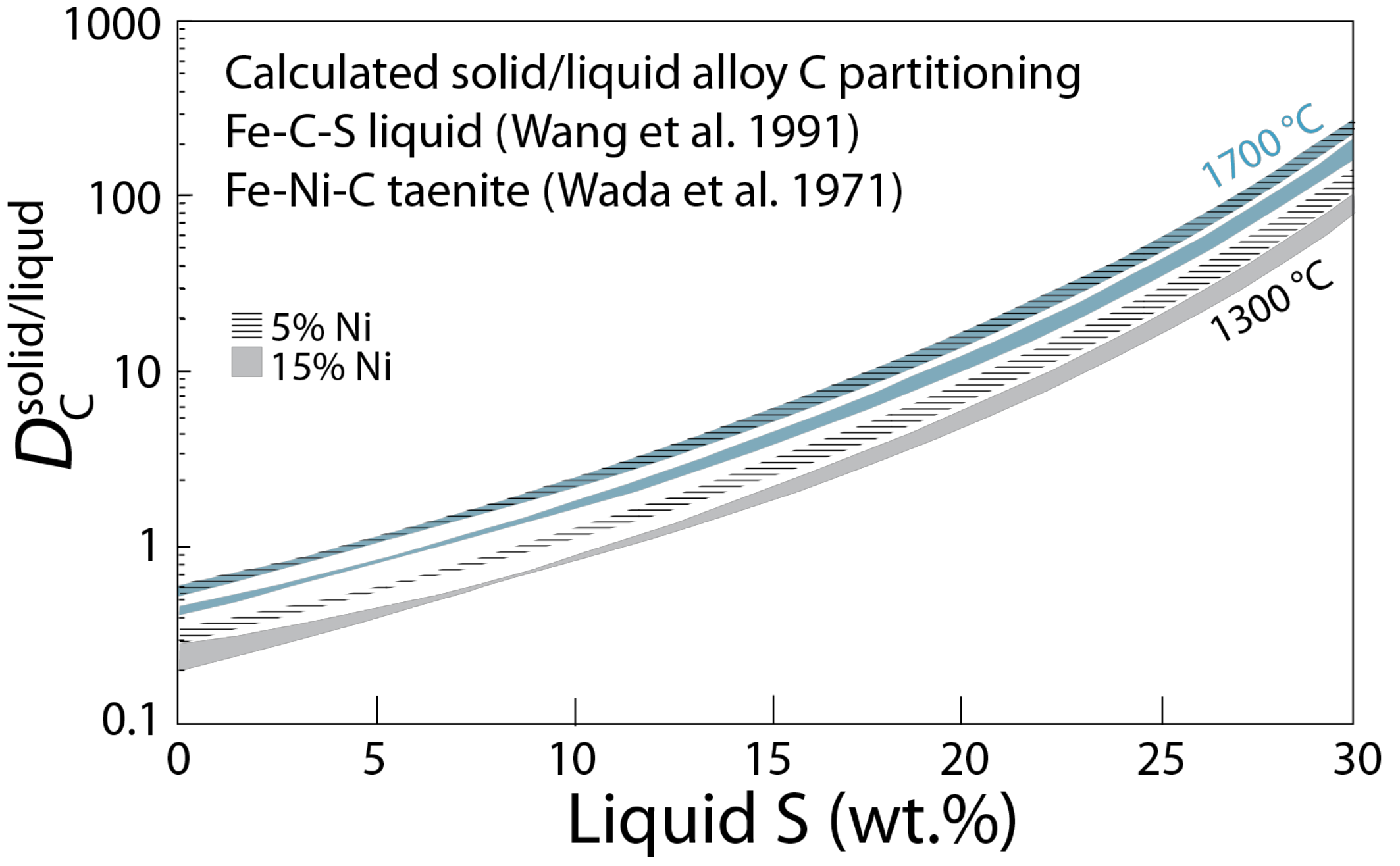

**Fig. S1** Calculated partitioning of C between Fe-Ni-C solid taenite (16) and Fe-C-S liquid (17) as a function of liquid S content at 1300 and 1700 °C and 5 and 15 wt.% Ni in the solid. Each band is calculated for solid C concentrations ranging from 0.01 to 0.5 wt.%. For calculations, see Supplementary File HirschmannPNAS-CFig2+S1Calc.xlsx

**Estimating bulk C in parent body cores.**

The procedure employed in this work to calculate the C of parent body cores assumes that the C concentration of iron meteorites represents that of a pure cumulate taenite, meaning solids precipitated from molten alloy. If these meteorites solidified instead as cumulate taenite combined with some fraction of solidified liquid trapped in the cumulate interstices, this assumption will not be wholly accurate. If that trapped liquid was enriched in C relative to the taenite ($D_C < 1$, applicable to liquids with less than ~9 wt.% S, Fig. S1, or 12 of the 18 groups shown in Figs. 2-4), then the actual aggregate solid would have had greater C than a pure cumulate, and the calculation would overestimate the parent liquid C content. This is the basis for the statement in the main text that the calculations provide maximum C contents of the iron parent body liquid.

If the trapped liquid had similar C relative to the taenite, as would be the case for parent bodies with liquid S contents 9-12 wt.% (4 of the 18 groups shown in Figs. 2-4), then the applicable $D_C$ would be close to unity and the trapped liquid would have little effect on the calculations. In the case where the trapped liquid was significantly enriched in S, as for the IIAB group if the S content was as inferred by the method of Chabot (20) (17 wt.% S, Fig. 2, Table S2) and the IC group (19 wt.%, Fig. 2, Table S2), then the applicable $D_C$ would be < 1 and the trapped liquid low in C. In this case, the bulk meteorite C content, and the calculated liquid C content could both be underestimates by a maximum factor of (1-*F*), where *F* is the fraction of a trapped liquid with negligible C. For example, for the IC group the estimated parent liquid of 0.05 wt.% (Fig. 2) would be revised upward to 0.063 wt.% if the analyzed samples were formed from 80% cumulate taenite and 20% trapped liquid (F=0.2). This difference is small compared to other uncertainties in estimates of C concentrations.

**Graphite saturation, alloy unmixing, and ureilite-like planetesimals**

As shown in Fig. 4, C-enriched core compositions produced from models of closed system planetesimal differentiation plot within the stability fields of Fe-C-S immiscibility and/or graphite saturation. This occurs for compositions derived from undepleted chondrites, but also for some calculated compositions derived from the depleted VDC planetesimal composition. The possibility of alloy unmixing or graphite saturation during differentiation of primitive parent bodies has several ramifications for planetesimal differentiation.

Fig. S2 shows selected tie-lines of coexisting liquids in the field of unmixing for the system Fe-C-S. (Note that the corresponding immiscibility field for Ni-containing liquid would be smaller, Fig.2). A particular bulk composition plotting along a given curve will unmix to that curve's endpoints. Bulk alloy compositions produced by core segregation of volatile-enriched planetesimals would yield graphite-saturated liquids whereas unmixed liquids formed from less volatile-enriched parents would be graphite-undersaturated.

If bulk alloy compositions derived from volatile-enriched planetesimals unmix, most of the mass becomes a C-rich, S-poor liquid, and a smaller fraction a C-poor, S-rich liquid. This is owing to mass balance, as the bulk composition is closer to the former rather than the latter. The lack of evidence for parent cores similar to such C-rich liquids suggests that volatile-enriched differentiated planetesimals are rare or absent. The less-abundant C-poor conjugate liquids would have > 0.1 wt.% C, which though not highly enriched in C, would be considerably greater C than compositions of medium C/S or low C/S parent body cores (Fig. S2). Such liquids would also be highly enriched (22-35 wt.%) in S. Thus, although unmixing could produce C-poor portions of cores formed from putative volatile-enriched planetesimals, this alone seems an inadequate explanation for inferred iron parent body compositions. Additional volatile-loss processes would be required.

Because all calculated iron parent body compositions plot away from the field of liquid immiscibility (Figs. 2, S2), we infer that bulk parent cores did not originate saturated with respect to a second liquid or graphite, and that early formation of stratified planetesimal cores based on Fe-S-C immiscibility, as suggested previously (21), is unlikely, although disequilibrium layered core segregation (22) cannot be excluded. Also, core segregation could be promoted as liquids evolve to more S-rich differentiates (21, 23, 24), though enrichment in residual liquid Ni inhibits immiscibility (18, 19) (Fig. 2) and could suppress separation of a second liquid.

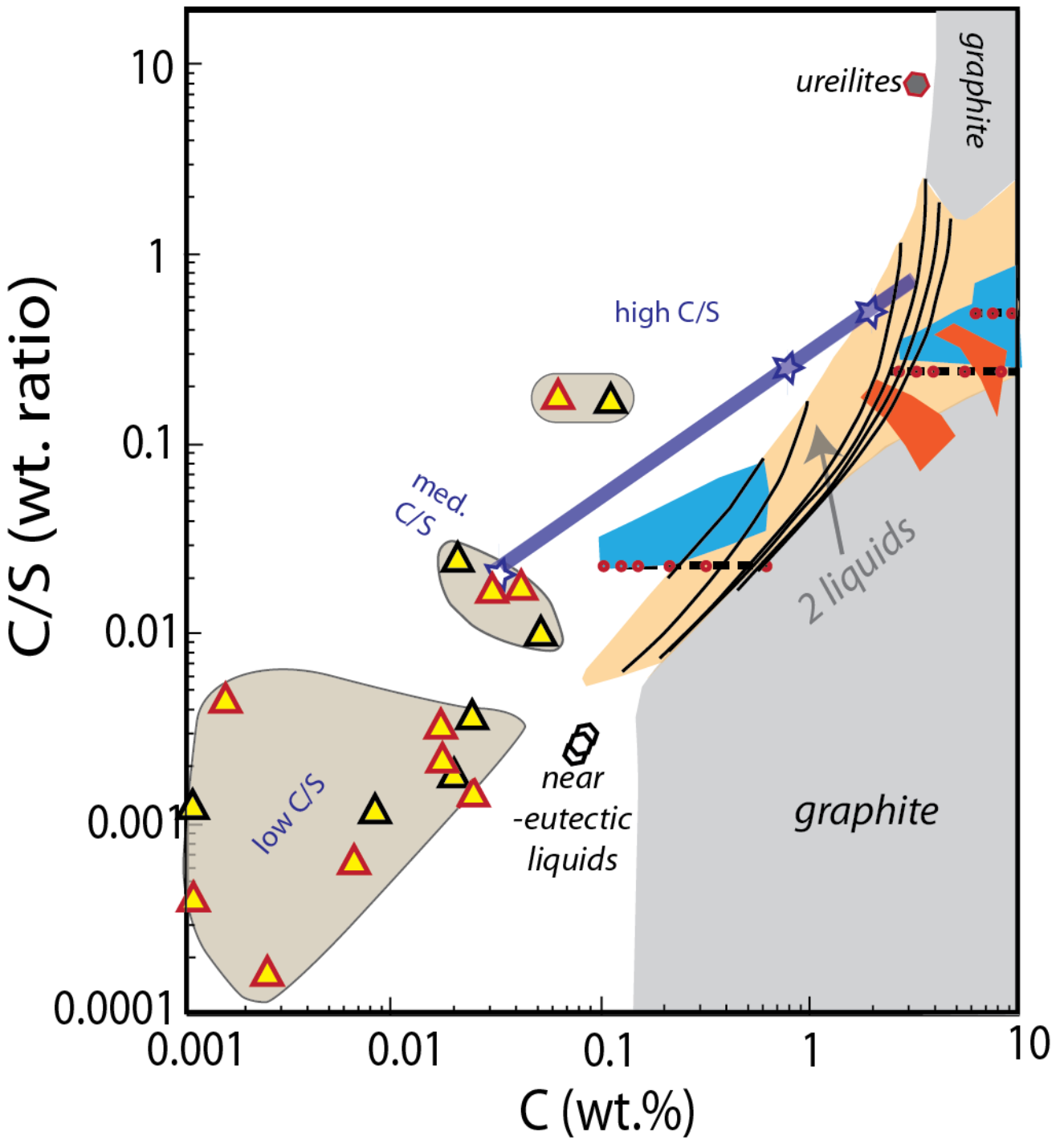

**Fig. S2** Calculated liquid alloy compositions formed during core formation of undepleted or less-depleted planetesimals (e.g., VRC, PDC; blue and red shaded regions and black dashed lines are calculated bulk core compositions, for guide see Fig. 4A.), which have bulk compositions within the 2-liquid stability field and will unmix to conjugate C-rich and S-rich liquids, as shown by the black tie lines, calculated from the model of Tafwidli and Kang (25). Tie lines are curved owing to the logarithmic axes. S-rich conjugate liquids have >0.1 wt.% C, and so are more C-enriched than inferred iron meteorite parent bodies. Ureilites (Table S1) are C-rich with high C/S, owing to accumulation of graphitic carbon (26). Polygons marked "near-eutectic liquids" are graphite and iron saturated S-rich melts at 1025-1125°C (27).

Ureilites are ultramafic achondrites that have lost a metal component (26) yet remain markedly enriched in C and poor in S, resulting in super-chondritic C/S (Table S1; Fig. S2). The substantial enrichments in C and C/S seen in ureilites are in some respects complementary to the low C and C/S of many parent body cores, which suggests an alternative scenario for the origin of C-poor planetesimal cores, discussed below. We note that ureilite genesis is known to be quite complex, involving multiple stages and mixing of diverse components (26, 28), and therefore emphasize that this hypothesis applies to bodies with similarities to ureilites – namely an ultramafic silicate planetesimal mantle containing graphite and residual alloy phases – and would not necessarily account for all the features of the actual ureilite parent body

Differentiation of a comparatively volatile-rich planetesimal (similar to the PDC or VRC compositions, Table S1) could lead to formation of two graphite-saturated liquids, rich in carbon and sulfur respectively. Draining of these liquids could produce a planetesimal core with some similarities to parent cores of iron meteorites, even though the precursor planetesimal was not markedly volatile-depleted, and would leave behind a silicate residue containing graphite and some evidence of residual S-rich and metal rich liquids, as observed in ureilites (29). However, this alternative hypothesis would need to address the challenges noted in the previous paragraphs. Namely, the S-rich liquid produced by unmixing would have > 22 wt. % S and significantly greater C than known parent body cores (Fig. S2). Any contribution of the C-rich conjugate liquid into the accumulated core would dilute the S, but greatly increase the C.

A more extreme version of this hypothesis is formation of metallic cores by extraction of a sulfur-rich liquid near the troilite-alloy eutectic at low temperatures (~1025-1125°C), leaving behind residual iron alloy. (Note that these temperatures are 100-200 °C lower than those inferred to have incited melting and smelting in the ureilite parent body (27, 30)). In graphite-saturated experiments at these conditions, Hayden et al. (27) produced liquids with ~0.08 wt.% C and ~32 wt.% S (Fig. S2). Compared to C-poor immiscible liquids discussed above, these low C contents are closer to inferred iron parent body compositions (Fig. S2), perhaps owing to the lower temperatures, but again, still too C-rich. Also, the S enrichments are extreme compared to those inferred for parent body cores. Finally, such a process would leave most of the highly siderophile elements in residual solid metal, and therefore in the planetesimal mantle, owing to strong solid/liquid partitioning (27). Though these processes are worthy of further investigation, we conclude that the iron meteorites are better explained by devolatilized planetesimals, rather than by retention of C in planetesimal mantles.

**Planetesimal core formation calculations**

Figure 4 depicts the calculated C and S concentrations of planetesimal cores calculated for four different models. Here we describe the calculations in greater detail, with each model denoted by the labels for the respective cartoons in Fig. 4. For each model, three different beginning planetesimal bulk compositions are considered, as illustrated in Fig.1 and given in Table S1, "VRC" (volatile-rich chondrite), "PDC" (partially devolatilized chondrite), and "VDC" (volatile-depleted chondrite).

Model A1 represents removal of C and S to the core in molten alloy, with no silicate melting, and has been calculated for planetesimals with weight fractions of alloy, $x_{\mathrm{alloy}}$, ranging between 5 and 30%. C and S are assumed to be removed quantitatively from the silicate, resulting in core concentrations that are $C_0/x_{\mathrm{alloy}}$, where $C_0$ is the concentration of C or S in the original bulk planetesimal.

Model A2 represents removal to the core of C and S from a molten silicate planetesimal interior that has a solid carapace preventing loss of volatiles to the surface or to space. The mass of the solid carapace is assumed to be negligible. The concentration of each element $i$ in the core, $C_i^{\mathrm{alloy}}$, is given by

$$C_i^{\text{alloy}} = \frac{C_i^0 D_i^{\text{alloy/silicate}} Q}{[(1-x_{alloy}) + x_{\text{alloy}} DQ]}\,, \qquad \text{Eqn. S1}$$

where $D_i^{\text{alloy/silicate}}$ is the applicable partition coefficient between molten alloy and silicate, and $Q$ is the alloy/silicate equilibration factor (31) ranging from 0 (no chemical equilibrium) to 1 (complete equilibrium). For each planetesimal bulk composition in Fig. 4A, the shaded region represents calculations from Eqn. S1 with values of $x_{\text{alloy}}$ ranging from 0.05 up to 0.3 and values of $Q$ ranging from 0.5 to 1. Values of $D^{\text{alloy/silicate}}$ depend on oxygen fugacity ($f_{O2}$) and alloy liquid composition and are given in Table S3.

Model *B* represents removal to the core of C and S from a molten silicate planetesimal interior that has experienced degassing to the surface either prior to or synchronous with metal-silicate equilibration. This calculation occurs in two steps, first volatile loss, followed by core separation, with the latter calculated from Eqn. S1 with an appropriately modified bulk composition. Volatile loss is considered to amount to 0-90% of the original planetesimal carbon, with no sulfur loss ($\Delta$S=0, forming the diagonal trend in the inset to Fig. 4B), carbon and sulfur loss in equal proportions ($\Delta$C=$\Delta$S, forming the horizontal trend), or intermediate relative losses of carbon and sulfur (forming the shaded area between the diagonal and horizontal trends).

Model *C* represents removal to the core of C and S from an entirely molten planetesimal in which core-forming alloy, molten silicate, and overlying degassed atmosphere equilibrate with one another following the model of Hirschmann (32). The atmosphere exerts vapor pressures on the underlying condensed liquids according to its mass and to planetesimal gravity, which is proportional to planetesimal radius. It is assumed that this atmosphere is later lost to space, but for simplicity, only after separation of alloy to the core. If part of the atmosphere is lost prior to silicate-alloy separation, then the resulting core would be further depleted in C and S owing to enhanced degassing. Partitioning between the three reservoirs, molten core, molten silicate, and atmosphere produces gives the mass fraction of element *i* in the atmosphere, $\frac{M_i^{\text{atm}}}{M_i}$, as.

$$\frac{M_i^{\text{atm}}}{M_i} = \left[\left(\frac{4}{9} G s_i r_i \rho R^2\right)\left((1-x_{\text{alloy}}) + x_{\text{alloy}} Q D_i^{\text{alloy/silicate}}\right) + 1\right]^{-1} \qquad \text{(Eqn. S2)}$$

(32), where G is the gravitational constant, $s_i$ is the solubility factor specifying the concentration of element *i* dissolved in silicate melt as a function of its partial pressure in the atmosphere (Table S3), ρ is the mean density of the planetesimal (3500 kg/m$^3$), and *R* is the planetesimal radius. The parameter $r_i$ is a mass factor that corrects for the difference between the mass of element *i* in the atmosphere and the mass of its gaseous species. Sulfur is assumed to degas as $H_2S$, and so

$r_S$=(38/36), and C as CO, giving $r_C$=(28/12). The corresponding core concentrations are given by

$$C_{\mathrm{i}}^{\mathrm{alloy}} = \frac{\left(1-\frac{M_{\mathrm{i}}^{\mathrm{atm}}}{M_i}\right) D_{\mathrm{i}}^{\mathrm{alloy/silicate}} Q}{[\left(1-x_{alloy}\right)+x_{\mathrm{alloy}} DQ]} . \qquad \text{(Eqn. S3)}$$

For the calculations shown in Fig. 4C, planetesimal radius is varied from 10 to 500 km, with larger radii resulting in greater retention of volatiles in the core. $x^{\mathrm{alloy}}$ is assumed to be 0.2 and $Q$ equal to unity, but resulting core compositions are not strongly sensitive to these assumptions.

**Table S1**
**C and S in Chondrites, Achondrites & Model Compositions**

| | S (wt.%) | C (wt.%) | C/S | Source |
|---|---|---|---|---|
| CI1 | 5.9 | 3.2 | 0.54 | (33) |
| CM2 | 3.3 | 2.2 | 0.67 | (33) |
| T. Lake2 | 3.65 | 5.81 | 1.59 | (34, 35) |
| CR2 | 0.87 | 2 | 2.29 | (36, 37) |
| CO3 | 2 | 0.45 | 0.23 | (33) |
| CV3 | 2.2 | 0.56 | 0.25 | (33) |
| CB3 | 2.19 | 0.28 | 0.13 | (38) |
| CK5 | 1.46 | 0.07 | 0.048 | (36, 37, 39) |
| CK6 | 0.98 | 0.03 | 0.034 | (38) |
| Avg. H | 0.11 | 2 | 0.055 | (33) |
| Avg. LL | 0.09 | 2.2 | 0.041 | (33) |
| Avg. L | 0.12 | 2.3 | 0.052 | (33) |
| EH | 0.4 | 5.8 | 0.069 | (33) |
| EL | 0.36 | 3.3 | 0.11 | (33) |
| H3 | 1.90 | 0.27 | 0.14 | (38) |
| H4 | 1.88 | 0.11 | 0.06 | (38) |
| H5 | 1.95 | 0.08 | 0.043 | (38) |
| H6 | 1.84 | 0.078 | 0.042 | (38) |
| L3 | 1.97 | 0.53 | 0.27 | (38) |
| L4 | 2.21 | 0.20 | 0.091 | (38) |
| L5 | 2.02 | 0.13 | 0.066 | (38) |
| L6 | 2.07 | 0.11 | 0.055 | (38) |
| LL3 | 2.27 | 0.31 | 0.13 | (38) |
| LL4 | 1.72 | 0.04 | 0.023 | (38) |
| LL5 | 1.85 | 0.07 | 0.038 | (38) |
| LL6 | 1.76 | 0.028 | 0.016 | (38) |
| Ureilites | 0.34 | 3.20 | 10.4 | (40-42) |
| Winonaites | 6.2 | 0.34 | 0.056 | (38, 40, 43) |
| Acapulcoites | 2.9 | 0.04 | 0.014 | (43, 44) |
| ***Model Compositions*** | | | | |
| DVC | 1.45 | 0.032 | 0.022 | |
| VRC | 4 | 2 | 0.5 | |
| PDC | 3.2 | 0.8 | 0.25 | |

**Table S2**
**Iron Meteorite Parent Body Compositions**

| | **S wt.%** | **C† wt.%** | **Liquid C† wt.%** | **Liquid C/S** | |
|---|---|---|---|---|---|
| **“cc” parents*** | | | | | |
| IIC | 8 | 0.019 | **0.023** | 0.0030 | S (45): C: (1, 2) |
| IIDw | 0.7 | 0.03 | **0.11** | 0.15 | S: (22) C; (1, 2) |
| IIDcl | 6 | 0.03 | **0.051** | 0.008 | S:(20) (lo S estimate) C: (1, 2) |
| IIDch | 12 | 0.03 | **0.019** | 0.00160 | S:(20) (hi S estimate) C: (1, 2) |
| IVBg | 1 | 0.0003 | **0.0011** | 0.0011 | S: (20, 46) C:(4) |
| IVBm | 1 | 0.006 | **0.021** | 0.021 | S: (20, 46) C:(1, 2) |
| SBT | 7 | 0.006 | **0.0087** | 0.0012 | S:(7) C: (2) |
| | | | | | |
| **“nc” parents*** | | | | | |
| IC | 19 | 0.15 | **0.026** | 0.0014 | S (47) C: (3) |
| IIABw | 6 | 0.01 | **0.017** | 0.0028 | S: (48) C: (4) |
| IIABc | 17 | 0.01 | **0.0025** | 0.00015 | S: (20)) C: (4) |
| IIIABw | 2 | 0.01 | **0.0230** | 0.015 | S: (49)C: (1, 2, 4, 5) |
| IIIABc | 12 | 0.01 | **0.0064** | 0.00054 | S: (20) C: (1, 2, 4, 5) |
| IVAwm | 0.4 | 0.016 | **0.061** | 0.15 | S: (50) C: (1, 2) |
| IVAclm | 3 | 0.016 | **0.043** | 0.014 | S: (20)(lo estimate) C: (1, 2) |
| IVAchm | 9 | 0.016 | **0.017** | 0.0019 | S: (20) (hi estimate) C: (1, 2) |
| IVAwg | 0.4 | 0.0004 | **0.0015** | 0.0038 | S: (50) C: (4) |
| IVAclg | 3 | 0.0004 | **0.0011** | 0.00036 | S: (20) (lo estimate) C: (4) |
| IVAchg | 9 | 0.0004 | **0.00044** | 0.00005 | S: (20) (hi estimate) C: (4) |

†For C, data in the second column is inferred mean concentration of C in the iron parent body of each group, and data in the second column (“liquid C”), is the thermodynamically calculated composition of the iron parent body core, as described in the text and depicted in Fig. 2. *“cc” and “nc” irons are those with isotopic affinities to carbonaceous and non-carbonaceous chondrites, as defined by (51). For further explanations of C concentrations, see SI Text.

**Table S3**
**Alloy/silicate partition coefficients**
**and vapor solubility**

| | | | | |
|---|---|---|---|---|
| Conditions | Alloy/silicate Partition coefficient | | Solubility Constant | |
| | $D_i^{alloy/silicate}$ | | $s_i$ (kg/Pa) | |
| | Carbon | Sulfur | Carbon | Sulfur |
| $f_{O2}$=IW-1 | 140 | 35 | 1 X $10^{-12}$ | 5 X $10^{-9}$ |
| $f_{O2}$=IW-2 | 550 | 27 | 5 X $10^{-13}$ | 5 X $10^{-9}$ |
| $f_{O2}$=IW-3 | 2200 | 13 | 2 X $10^{-13}$ | 5 X $10^{-9}$ |
| S-rich | 10 | 100 | 5 X $10^{-13}$ | 5 X $10^{-9}$ |

Values are applicable to partitioning between ultramafic silicate melt and Fe-C-S melts at oxygen fugacities ($f_{O2}$) 1, 2, and 3 log units below that defined by the coexistence of iron-wüstite (IW). The fourth set of values is applicable to S-rich (~20 wt.% S) alloy melts (52). All values from Hirschmann (32) except for alloy-silicate partition coefficients for sulfur-rich melts (52).

## Additional References